\documentclass[12pt]{iopart} 
\pdfoutput=1
\usepackage{graphicx}
\usepackage{color}
\usepackage[english]{babel}
\usepackage{epsfig}
\usepackage{amssymb}
\usepackage{ulem}
\usepackage{color}
\newcommand{\DegreeC}{$^\circ$C}
\begin{document}
\title{Nanostructuring Graphene by Dense Electronic Excitation}

\author{O.~Ochedowski$^1$, O.~Lehtinen$^2$, U.~Kaiser$^2$, A.~Turchanin$^3$, B.~Ban-d`Etat$^4$, H.~Lebius$^4$, M.~Karlusic$^5$, M.~Jaksic$^5$, and M.~Schleberger}
\address{$^1$Fakult{\"a}t f{\"u}r Physik and CENIDE, Universit{\"a}t Duisburg-Essen, 47048 Duisburg, Germany}
\address{$^2$Universit{\"a}t Ulm, Electron Microscopy Group of Materials Science, 89081 Germany}
\address{$^3$Friedrich Schiller University Jena, Institute of Physical Chemistry, 07743 Jena, Germany}
\address{$^4$CIMAP, (CEA-CNRS-ENSICAEN-UCN), blvd Henri Becquerel, 14070 Caen, France}
\address{$^5$Ruder Boskovic Institute, Bijenicka cesta 54, 10000 Zagreb, Croatia}
\author{$^1$M.~Schleberger}


\begin{abstract}
The ability to manufacture tailored graphene nanostructures is a key factor to fully exploit its enormous technological potential. We have investigated nanostructures created in graphene by swift heavy ion induced folding. For our experiments, single layers of graphene exfoliated on various substrates and freestanding graphene have been irradiated and analyzed by atomic force and high resolution transmission electron microscopy as well as Raman spectroscopy. We show that the dense electronic excitation in the wake of the traversing ion yields characteristic nanostructures each of which may be fabricated by choosing the proper irradiation conditions. These nanostructures include unique morphologies such as closed bilayer edges with a given chirality or nanopores within supported as well as freestanding graphene. The length and orientation of the nanopore, and thus of the associated closed bilayer edge, may be simply controlled by the direction of the incoming ion beam. In freestanding graphene, swift heavy ion irradiation induces extremely small openings, offering the possibility to perforate graphene membranes in a controlled way.
\end{abstract}

\maketitle


\maketitle

\noindent
{\bf Introduction}\\
Folding is an old an every-day technique which can be used to shape a material but also to modify material properties such as mechanical strength. While folding is usually applied in the macroscopic world, recent work demonstrated that it might be useful also for nanoscaled materials such as graphene \cite{Blees.2015}, the most prominent of the two-dimensional materials. In the case of graphene, folding has attracted quite a lot of attention in particular, as it offers the chance to tune its mechanical, chemical, and electronic properties \cite{Wu.2013,Hallam.2014}. For example, graphene folds locally strengthen the material \cite{Zheng.2011}, they may be used as transport channels, and graphene origami boxes could serve as containers for hydrogen storage \cite{Zhu.2014} for which then an electric field can be used to unfold graphene and release the stored gas. At the point where folded graphene is bent, a closed bilayer edge is formed. In these closed bilayer edges the covalent bonds are bent, which effectively changes the local chemical reactivity and transport properties \cite{Ortolani.2012}. As a consequence a strong magneto-photoelectric effect \cite{Queisser.2013}, bandgap openings \cite{Feng.2009,Zhan.2011}, and enhanced spin-orbit interaction \cite{Costa.2013} are expected. Furthermore, folding of graphene is accompanied by the formation of bilayer graphene, either in the regular stacking or in its twisted form. The latter shows rotational angle dependent optical properties \cite{Campos.2013}, flat bands close to the Fermi energy \cite{SuarezMorell.2010}, chiral tunneling \cite{He.2013}, and large interlayer resistivity, strongly dependent on temperature \cite{Kim.2013}. 

This huge scientific and technological potential of folded graphene may however only be exploited if reliable methods for controlled folding are at hand. Folding of graphene has so far been achieved by ultrasound sonication \cite{Zhang.2010}, the use of chemicals \cite{Allen.2009}, by using the tip of an atomic force microscope (AFM) \cite{Liu.2006,Temmen.2013}, femto-second laser ablation \cite{Yoo.2012}, and swift heavy ion irradiation (SHI) \cite{Akcoltekin.2011,Ochedowskii.2014}. Most the aforementioned methods create foldings in a more or less random way with respect to size and orientation, as well as the number of foldings. In addition, they require edges, i.e.~folding of graphene without a pre-existing defect cannot be achieved. The irradiation with swift heavy ions is a notable exception and may in principle overcome all these disadvantages, because type of ion, energy, angle of incidence and fluence are easy to control parameters, and in addition, SHI may induce foldings anywhere in a graphene flake. In general, this type of ion beams represent a powerful processing tool to produce nanoscaled material modifications which are unachievable by other methods \cite{Akcoltekin.2007,Lang.2009,Ridgway.2011,Ridgway.2013,Ochedowski.2013b,Ochedowski.2014,Papaleo.2015}. Due to their high kinetic energy, typically several hundred keV per nucleon and above, the main interaction with matter is not via nuclear collisions, but via dense electronic excitation. Thereby extreme energy densities are achieved and the affected area is centred around the trajectory of each ion. In a grazing incidence geometry the resulting material changes at the surface may extend up to several microns. Thus, nanostructuring of graphene by SHI seems perfectly feasible. However, there exists a profound lack of knowledge as so far neither the mechanism nor the potential of this approach have been investigated yet. 

To assess the potential and to determine the conditions for nanostructuring graphene, we have prepared and irradiated supported and freestanding graphene samples with swift heavy ions. We have analysed the resulting morphological changes with optical spectroscopy and various microscopy methods down to the atomic scale and determined the influence of the ion energy. Finally, we varied the experimental conditions to investigate the influence of the substrate and graphene itself on the irradiation-induced graphene nanostructures. This paper offers evidence that SHI irradiation can be used to perforate supported and even freestanding graphene. It provides the necessary threshold energy per track length for nanostructuring and the details on how to obtain certain types of graphene nanostructures. Thus, the full potential of SHI irradiation in conjunction with the novel material class of two-dimensional solids is revealed.\\

\begin{figure}[htb]
\centering
\includegraphics[width=14cm,keepaspectratio]{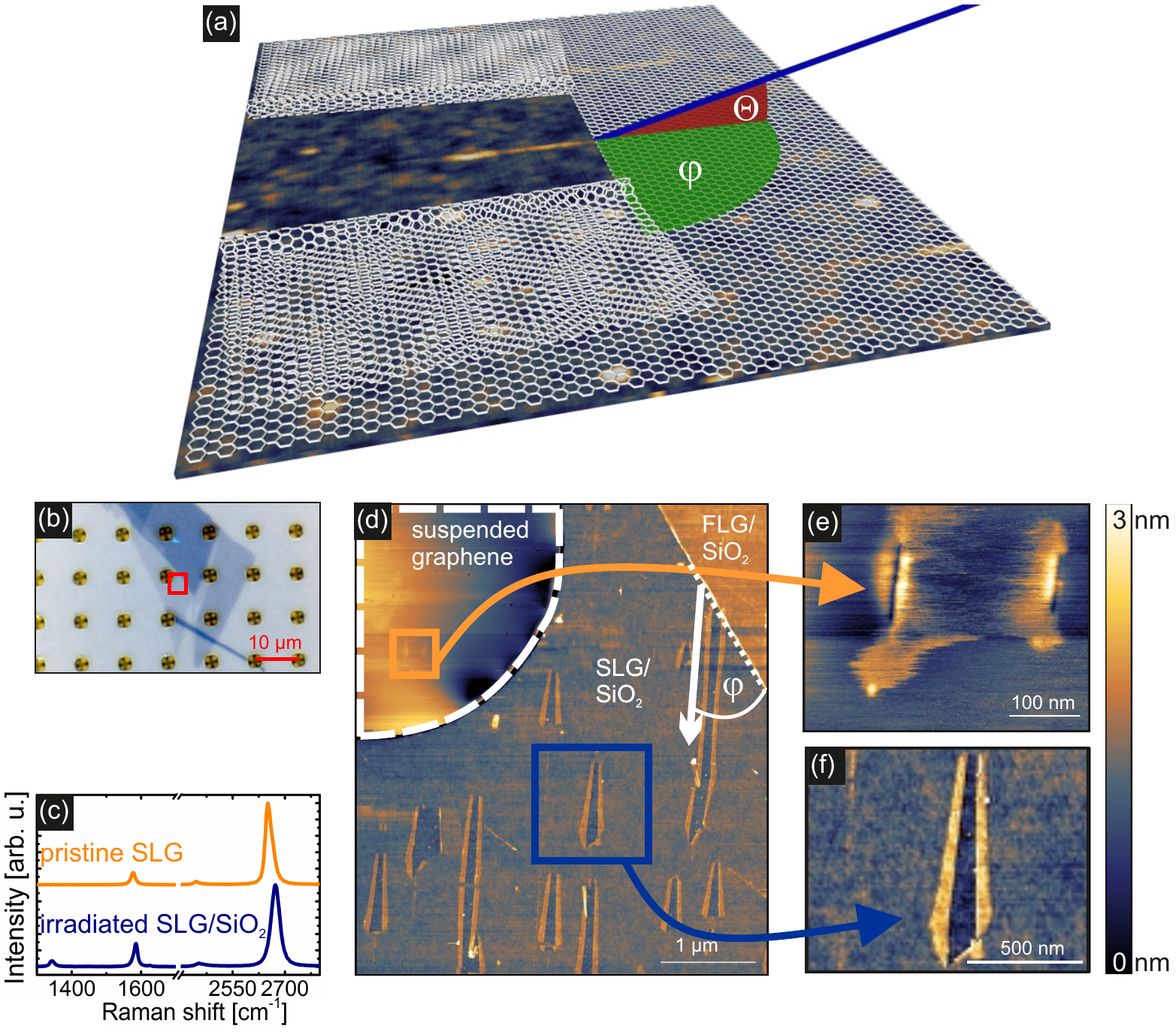}
\caption{Comparison between nanostructures obtained in substrate supported and suspended graphene after a typical experiment in glancing angle geometry (a). (b) Optical image of graphene exfoliated on a patterned SiO$_{2}$ substrate. (c) Raman spectra taken on the pristine (orange) and on the irradiated graphene sheet (blue), here with 91~MeV Xe, $\Theta=0.1^{\circ}$. A distinct $D$-band at 1350 cm$^{-1}$ is observed. (d) AFM topography of the irradiated graphene flake within the red box in the optical image; ion beam direction in (d) marked by the white arrow. Note, that the edge of the few layer graphene (FLG) in the upper right corner runs parallel to a low-indexed direction of the single layer underneath, which was used to determine $\varphi=33\pm3^{\circ}$ for this sample. (e) Nanoscaled slit-pore in the suspended graphene sheet. Their length is about one tenth of the length of (f) the opening found in the substrate supported graphene. Here, the folded bilayer is clearly seen.}
\label{1}
\end{figure}

\noindent
{\bf Results and discussion}\\
To investigate nanostructuring of graphene by SHI induced folding, the ion beam is directed towards the sample as schematically depicted in Fig.~\ref{1}(a). For a first experiment, graphene has been exfoliated on a pre-patterned SiO$_{2}$ substrate (3~$\mu$m diameter holes, 8~$\mu$m in depth written by photolithography and consecutive dry etching). A typical, high quality single layer (see Raman spectrum in Fig.~\ref{1}(c)) graphene flake on such a patterned substrate is shown in the optical image in Fig.~\ref{1}(b). The red box in Fig.~\ref{1}(b) marks the area where the AFM image has been taken which shows suspended graphene in the upper left corner and SiO$_{2}$ supported graphene elsewhere, see Fig.~\ref{1}(d). 

This sample has been irradiated with 91~MeV Xe ions at an angle of incidence $\Theta=0.1^{\circ}$ with respect to the surface. The irradiation induced defects show up as a notable $D$ peak in the Raman spectrum. Atomic force microscopy images taken from graphene supported by the SiO$_{2}$ substrate reveal that these defects are extended foldings which are in average one micron long (Fig.~\ref{1}(d,f)) and are created with an efficiency, i.e.~the number of foldings per ion, of 1. At first glance, no foldings or modifications at all can be observed in the suspended graphene sheet in Fig.~\ref{1}(d). However, the zoom-in in Fig.~\ref{1}(e) shows that nanoscaled slit pores appear in suspended graphene. The length of these slit pores is only about one tenth of the length of the foldings on the SiO$_{2}$ substrate. 

Fig.~\ref{1} presents the three characteristic irradiation induced graphene nanostructures, (twisted) bilayers, closed bilayer edges, and pores in supported graphene as well as nanopores in freestanding graphene. Note, that foldings and pores can be created directly within the flake and that no open edges are required as with other methods. We will proceed by discussing the conditions under which each of these structures might be fabricated and how these can be met. An important issue is certainly the minimum deposited energy per track length that is required to obtain a given modification. In the energy regime of swift heavy ions, typically projectiles with 0.5 to a few MeV per atomic mass unit, the deposited energy will depend on the kinetic energy of the ion and its mass. The deposited energy per path length $dE/dx$ is usually given in keV/nm and is called electronic stopping power $S_e$, because the primary energy transfer channel involves electronic excitations and ionization of the target material. The subsequent processes may involve non-thermal and thermal transitions such as ultrafast melting and Coulomb explosion on fs time scales~\cite{Ultrafast,Fleischer.1965}, and melting on the ps time scale due to electron phonon coupling (thermal spike)~\cite{Toulemonde.1992}. In order to determine the threshold for material modification to occur in terms of stopping power, graphene on SiO$_{2}$ samples have been irradiated with SHI of different stopping powers in the range from 2 -- 18~keV/nm (by choosing different ions at different kinetic energies, see Methods).

\begin{figure}[ht]
\centering
\includegraphics[width=15cm,keepaspectratio]{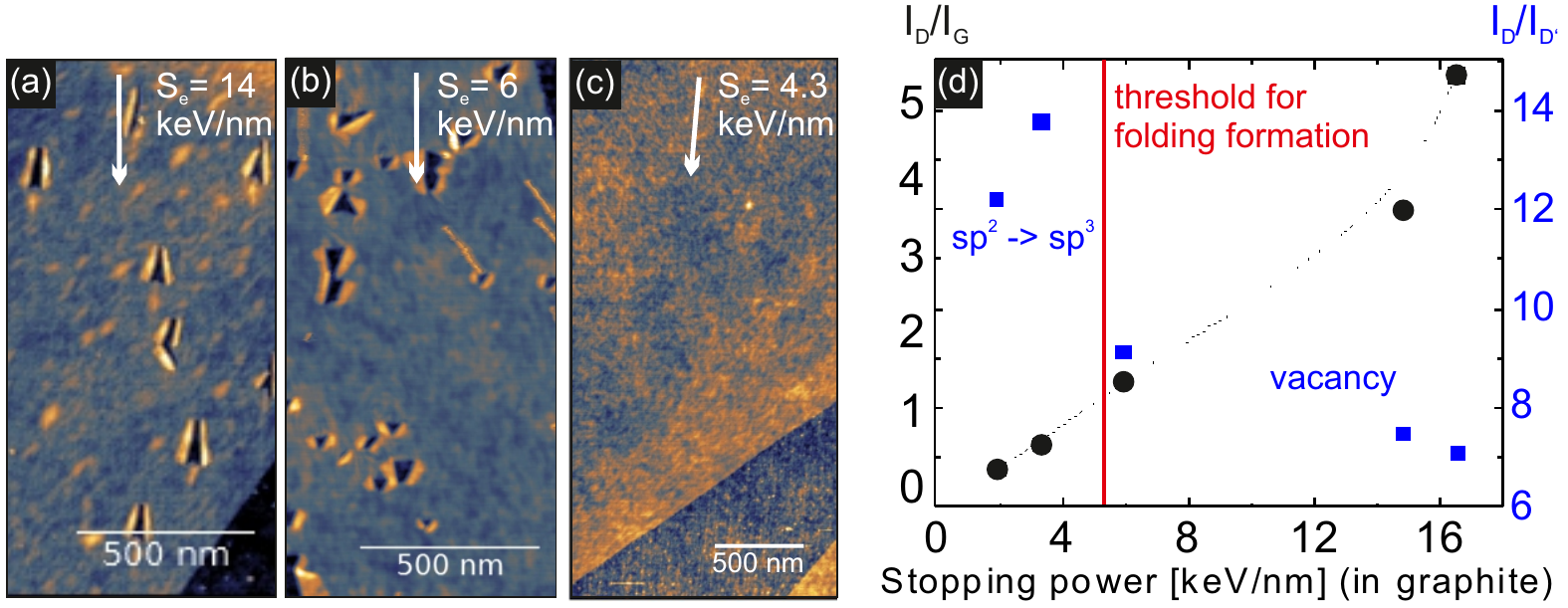}
\caption{Influence of the stopping power on the folding length. The angle of incidence was kept constant at $\Theta=1.8^{\circ}$. (a) AFM topography of SLG on SiO$_{2}$ irradiated with 84~MeV Ta ions yielding $dE/dx=14$~keV/nm, (b) irradiated with 23~MeV I yielding $dE/dx=6$~keV/nm, and (c) irradiated with 15~MeV Si yielding $dE/dx=4.3$~keV/nm. Height scales are 0 to 5~nm. (d) Analysis of $D/G$ (black dots) and $D/D'$ (blue squares) intensity ratios from Raman spectra. Black line to guide the eye. The red line represents the minimum stopping power that induces foldings detectable by AFM.}
\label{2}
\end{figure}

As an example we show in Fig.~\ref{2} graphene samples irradiated with 14 keV/nm, 6~keV/nm, and 4.3~keV/nm, all with the same nominal incidence angle of $\Theta=1.8^{\circ}$. Note, that we took great care to determine the accurate incidence angle in this experiment (see \cite{Ochedowski.2014} for details) so that any deviation between these samples due to a difference in the incidence angle $\Theta$ can be excluded. Comparing the AFM measurements in Fig.~\ref{2}(a)-(c) it can be clearly seen, that the size of foldings (and the corresponding pores, respectively) in graphene depends on the stopping power: While the foldings with 6~keV/nm are typically 60~$\pm$ 20~nm in length, foldings created with 14~keV/nm are twice as large (130 $\pm$ 40~nm). Note that the scatter in length maybe due to the statistical nature of the electronic stopping and is expected to be even larger in the vicinity of a threshold. Furthermore, Fig.~\ref{2}(c) shows that for lower stopping powers not only the size of the foldings is affected, but also the folding efficiency approaches zero. We found that the minimum stopping power down to which foldings can be observed is $5\pm1$~keV/nm, marked by the red solid line in Fig.~\ref{2}(d). This threshold is somewhat higher than the one recently reported for perpendicular irradiation \cite{WeisenLi.2015} and lower than the threshold for defect creation in graphite~\cite{Liu.2001} where foldings have never been observed. The surprisingly low value offers the technical advantage that also smaller accelerators can be used for folding of graphene.

The threshold for folding to occur can be compared to the threshold for defect creation as determined from Raman spectra taken at an excitation wave length of 532~nm from samples irradiated with ions of varying stopping powers but at a constant fluence, here 5$\cdot$10$^{4}$~ion/$\mu$m$^{2}$ under perpendicular incidence. The plot in Fig.~\ref{2}(d) shows a steady increase of the I$_{D}$/I$_{G}$ ratio (left scale, black dots) with increasing ion stopping power. Apart from the I$_{D}$/I$_{G}$ ratio which quantitatively describes the size of the defects, the I$_{D}$/I$_{D'}$ ratio can be used to investigate the nature of defects \cite{Eckmann.2012}. Eckmann et al.~found a maximum for the I$_{D}$/I$_{D'}$ ratio ($\cong$~13) for defects due to $sp^{3}$ hybridization of carbon with other atoms. Lower I$_{D}$/I$_{D'}$ ratios are indicative of vacancy like ($\cong$~7) and boundary like defects ($\cong$~3.5). Note, that In Fig.~\ref{2}(d) above the threshold stopping power for foldings (red line) the I$_{D}$/I$_{D'}$ ratio (right scale, blue squares) is between 7 and 9 which corresponds to vacancy like defects. Below the threshold stopping power for foldings (as determined by AFM) an increase of the I$_{D}$/I$_{D'}$ ratio to 12 -- 14 can be observed which corresponds to $sp^{3}$ hybridization. This indicates that not only the size of the structurally-disordered area but also the nature of defects can be controlled by the SHI stopping power. Furthermore it points towards a number of aligned vacancy defects acting as the artificial point of failure necessary for folding, while $sp^{3}$ defects are not as effective. This is in full agreement with the reported mechanical properties of defective graphene. In terms of stiffness and intrinsic strength, $sp^{3}$ defective graphene shows only a slightly reduced breaking strength while a significant degradation of mechanical properties was observed in the vacancy regime \cite{Zandiatashbar.2014}. 
\\

\begin{figure}[ht]
\centering
\includegraphics[width=15cm,keepaspectratio]{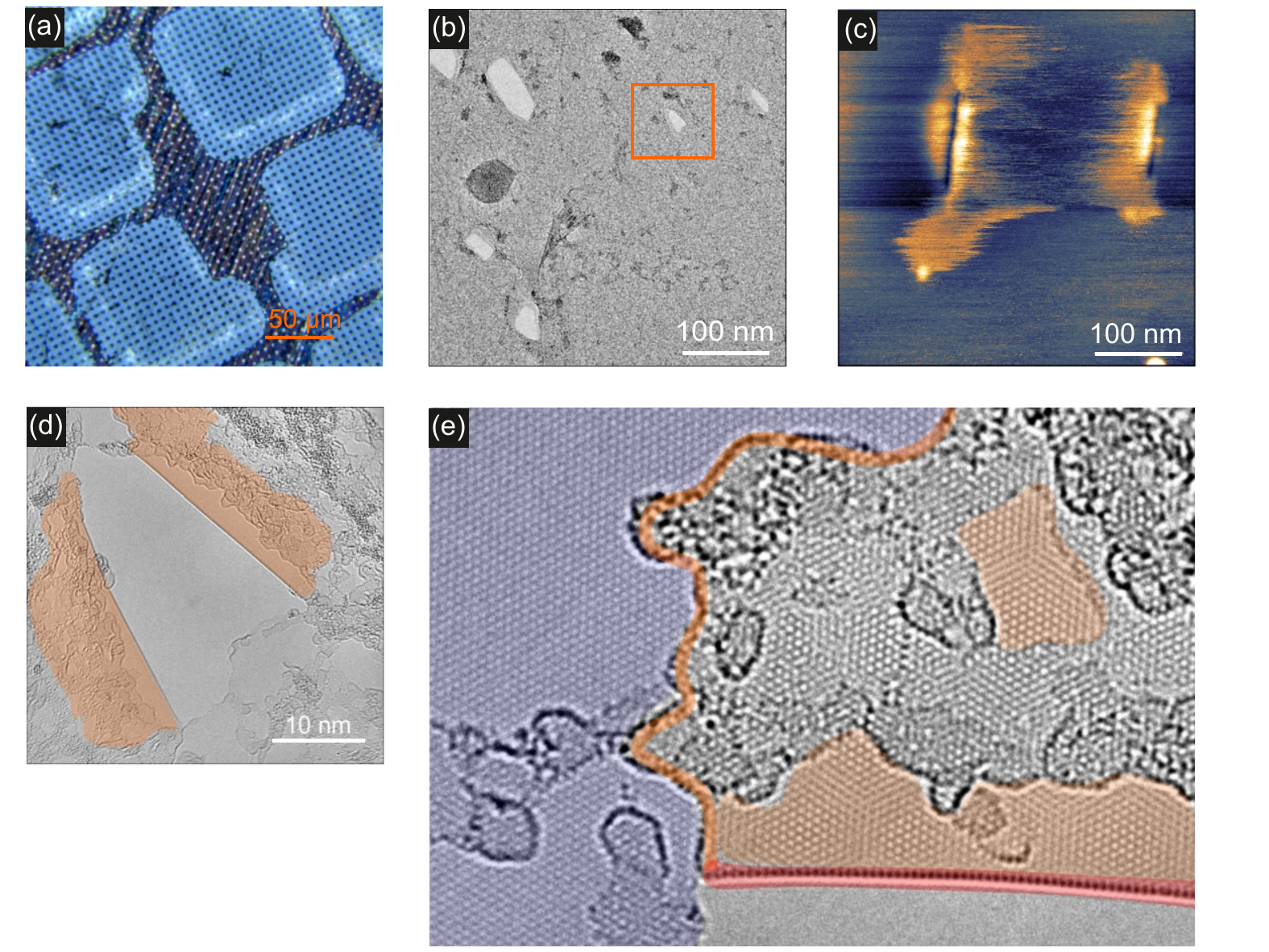}
\caption{SHI modifications in suspended graphene ($dE/dx=15$~keV/nm). (a) Optical image of a TEM grid onto which CVD graphene has been transferred. (b) AC-HRTEM measurement of the TEM-grid suspended graphene after irradiation with SHI. Here, contaminants can be seen as dark areas on the graphene flake which are due to the transfer process and adsorbates/hydrocarbons from the air \cite{AlgaraSiller.2014}.  For comparison we show again and in the same scale the AFM image from fig.1, where the typical morphology after grazing incidence irradiation of freestanding graphene is a slit pore. (d) Zoom-in of the AC-HRTEM measurement reveals that nanoscaled rifts from the AFM image are in fact foldings. (e) Atomically resolved image shows twisted bilayer (orange and gray) and closed bilayer egde (red).}
\label{3}
\end{figure}

In the following we will discuss in more detail the nanostructures presented in Fig.~\ref{1}, and how they may be fabricated in a controlled manner. We will begin with the nanoscaled slit pores in freestanding graphene. It is quite difficult to image suspended graphene with AFM in tapping mode as the oscillating tip in this operation mode comes into contact with the membrane which easily bends due to the applied force. This results in a blurred topography image offering only little details, see Fig.~\ref{1}(e). To circumvent this problem and study the ion induced modification with the highest possible resolution, we investigated the samples by aberration-corrected high-resolution transmission electron microscopy (AC-HRTEM). For the AC-HRTEM measurements, graphene grown on a copper foil was transferred using polymethylmethacrylate (PMMA) onto a Quantifoil-TEM grid which is shown in Fig.~\ref{3}(a). The suspended graphene sheets are located on top of the holes with a diameter of 2 -- 3~$\mu$m which is sufficiently large for the AC-HRTEM measurement. 

To compare the results from AFM (Fig.~\ref{1}(c)) with the AC-HRTEM measurements, both images are shown next to each other and using the same scale in Fig.~\ref{3}(b,c). Most notably, the slit pores seen in the AFM image are revealed by the AC-HRTEM measurement to be miniature foldings (Fig.~\ref{3}(d)), with an aspect ratio (length to width) of approximately 2 to 1. This zoom-in also reveals that surprisingly, not too many carbon atoms are missing as folding back the orange areas (where the bilayer could be roughly identified) would again close the nanopore. An atomically resolved image of such a folding clearly reveals the unique nature of the nanostructure created by the ion impact, see Fig.~\ref{3}(e). Here, the unaffected single-layer graphene area is colored in violet, while the border of the folded part is marked orange to outline the extension of the bilayer. The line along which graphene has been flipped over and the closed bilayer edge has formed, is highlighted in red. The high-resolution AC-HRTEM image reveals a straight edge at the semicircular bend with no missing carbon atoms, corroborating the hypothesis that the zipped-up graphene layer flips over and attaches itself to the unaffected material underneath (or above), eventually trapping residual atoms and contaminants in the process. Despite the disruptive nature of the process, the folded areas maintain to a large extent a well-ordered graphene structure, resulting in twisted bilayer sheets. This becomes evident from the distinct Moir\'e pattern (areas coloured in orange) which is due to an orientational lattice mismatch between the first and second layer graphene of $\sim 5^{\circ}$ in this case. Areas on the bilayer colored in gray are contaminated which might play a role for the unusual and differing heights typically found in AFM images. Finally, we wish to point out, that typically the major part of the boundary of a nanopore consists of an intact closed bilayer edge. Also, it is safe to assume that the aspect ratio of the nanopores can be controlled by the incidence angle as is the case for supported graphene. This fact, together with the unique and well-defined edge configuration, might prove advantageous for the fabrication of porous graphene membranes for filtering applications where pore size and the termination of the pore edge play an important role for filter performance~\cite{Liu.2014}. \\

If non-twisted, twisted bilayers, or closed bilayer edges are to be fabricated in supported graphene samples, the most important parameters will be the two angles introduced in Fig.~\ref{1}(a): The polar angle $\Theta$ which denotes the angle of incidence of the ion with respect to the sample surface and the azimuthal angle $\varphi$, the angle between the incident ion and a given crystallographic edge of the 2D-crystal (which may be either armchair or zigzag). Although irradiation under perpendicular incidence can be used to induce defects in graphene and graphene related materials, as has been shown for singly charged ions \cite{Lucchese.2010}, highly charged ions \cite{Hopster.2014,Ritter.2013} and swift heavy ions \cite{Ochedowski.2013}, no foldings are introduced under these conditions. However, under glancing incidence angles individual SHI do cause foldings of different length and shape, for which a two-step model has been proposed \cite{Akcoltekin.2011}. In the first step the SHI irradiation causes via electronic excitations an extended line of defects, e.g.~in form of a transition from hexagonal carbon rings to pentagonal and heptagonal ones. Subsequent to the creation of these primary defects, the hillocks which are formed by the SHI interaction with the substrate surface, are pushing through the graphene layer which is then unzipped along the ion trajectory. This hypothesis implies that the glancing incidence condition is a prerequisite for the folding as it yields a significant number of defects all aligned in a row to sufficiently weaken the graphene layer. Therefore, we have determined the minimum polar angle which still allows for a successful folding process by irradiating a set of SiO$_{2}$ supported graphene samples at a constant fluence of 15~ions/$\mu$m$^{2}$ and varying incidence angles $\Theta$.

\begin{figure}[ht]
\centering
\includegraphics[width=15.5cm,keepaspectratio]{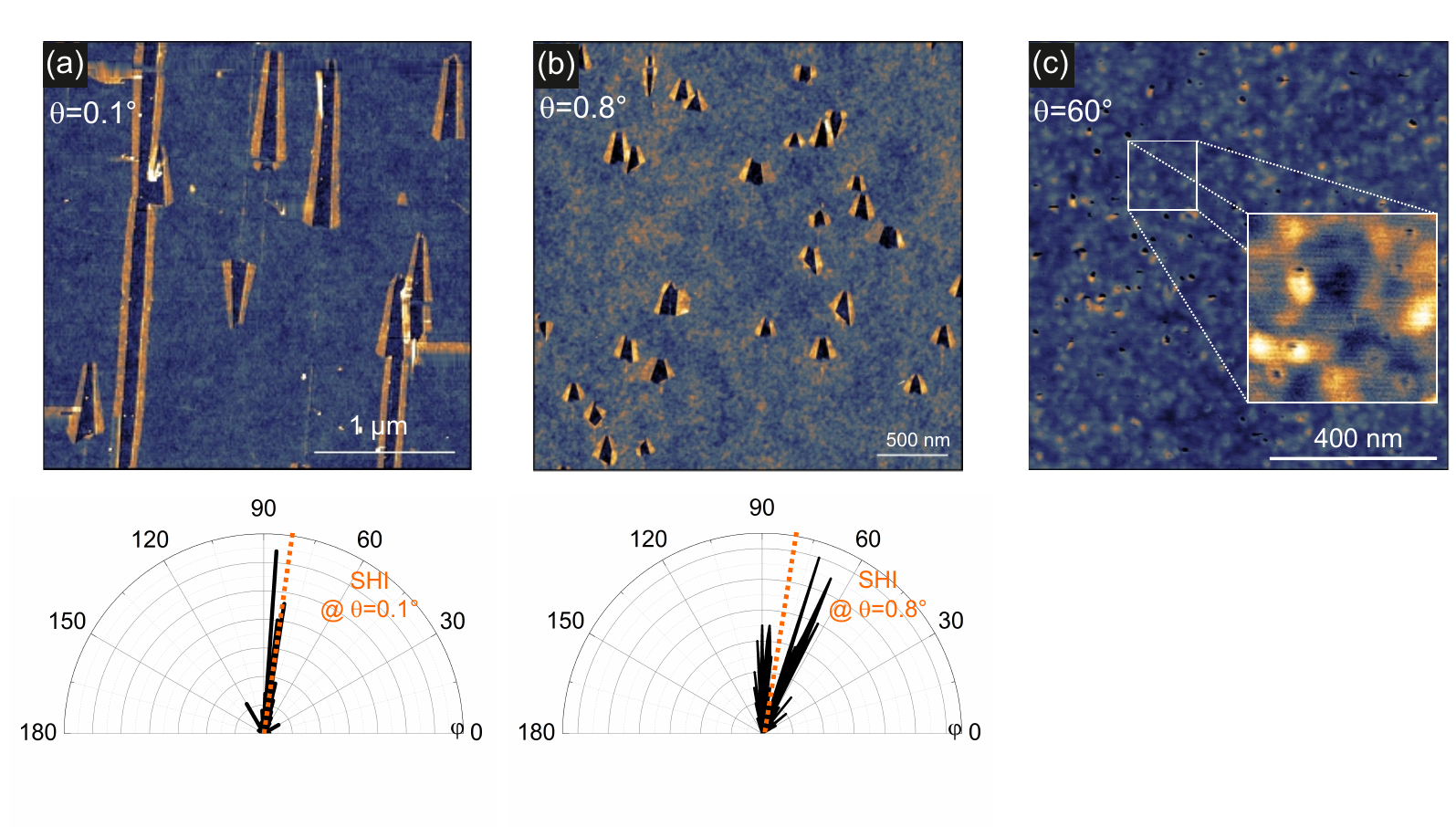}
\caption{Size of the foldings and orientation of the closed bilayer edge of graphene on SiO$_{2}$. (a) AFM image of foldings after irradiation with $dE/dx=15$~keV/nm at $\Theta=0.1^{\circ}$. Height scales are 0 to 5~nm. For this ultra grazing incidence angle, foldings are mainly oriented along the SHI trajectory (dashed orange line) as the corresponding polar diagram shows. The error of $\varphi$ is in the range of $\pm$ 3$^\circ$ mainly because of piezo drift during AFM measurement. (b) For less grazing incidence angle (here $\Theta=0.8^{\circ}$) foldings are shorter and are mainly oriented close to low-indexed directions of the graphene lattice. (c) At large incidence angles such as $\Theta=60^{\circ}$, nanopores with a diameter $\leq10$~nm are observed by AFM.}
\label{4}
\end{figure}

Exemplary results of this irradiation series shown in Fig.~\ref{4}(a)-(c), make it evident that extremely long (in the range of microns) closed bilayer edges can be fabricated, while the width of the associated bilayer is on the order of a few 10 nanometres and thus close to what would typically be achieved with lithographic techniques~\cite{Yu.2013}. At large angles of incidence, e.g.~$\Theta=60^{\circ}$ as shown in Fig.~\ref{4}(c), 
the irradiation does not yield detectable foldings anymore, instead sub-10~nm pores are observed. Detectable folding by means of ambient AFM occurs in graphene on SiO$_2$ up to incidence angles of $\Theta\simeq 45^{\circ}$. For comparison we like to point out that irradiation of graphene on Ir(111) with keV ions at comparable angles of incidence yields chains of vacancy clusters~\cite{Standop.2013}. In that energy regime, the primary energy dissipation takes place via atomic collisions and thus, the occurrence of nanopores framed by foldings is a unique consequence of the dense electronic excitations triggered by the SHI. 

Controlling the length of graphene nanostructures with closed bilayer edges is only one factor. To actually make use of twisted bilayer and closed bilayer edges, with their specific electronic and chemical properties, it is absolutely necessary to control the lattice mismatch or the orientation of the closed bilayer edge, respectively. It is well known that in mechanically exfoliated graphene samples the angles between edges are typically given in multiples of 30$^{\circ}$. This indicates that graphene is cleaved preferably along its low-indexed crystalline directions and either zigzag or armchair edges are generated \cite{Ni.2009}. For the same reason, artificially induced foldings of graphene tend to exhibit folding angles close to 30$^{\circ}$ as well \cite{Zhang.2010}. With SHI irradiation this limitation can be overcome. To demonstrate this, we have analyzed AFM images such as Figs.~\ref{4}(a),(b) by measuring the azimuthal angle $\varphi$ of the closed bilayer edge with respect to a given single layer graphene edge of the same flake which may be either armchair or zigzag. In Fig.~\ref{4}(a),(b) we show the corresponding angle distributions determined for two different incidence angles $\Theta$. For an ultra glancing incidence angle like $\Theta$=0.1$^{\circ}$, the orientation of the closed bilayer edges does not show the characteristic multiples of 30$^{\circ}$, instead $\varphi$ is almost exactly oriented along the direction of the impinging SHI as marked by the dashed orange line in the polar diagram of Fig.~\ref{4}(a). In Fig.~\ref{4}(b) the incidence angle was increased to $\Theta$=0.8$^{\circ}$ and the resulting closed bilayer edges show a $\varphi$-distribution which is no longer aligned with the direction of the impinging SHI, but is centered around the 60$^{\circ}$ and 90$^{\circ}$ directions, corresponding to the graphene zigzag and armchair orientations (or vice versa). Going to even higher incidence angles results in a widespread distribution of $\varphi$ with no obvious correlation to the ion beam direction (data not shown). In conclusion, it can be said that we are able to obtain graphene nanostructures with closed bilayer edges oriented exactly along the SHI trajectory with an efficiency of more than 90\% at low incidence angles like $\Theta=0.1^{\circ}$. Only for higher incidence angles, the orientation of the closed bilayer edge typically exhibits multiples of 30$^{\circ}$, i.e.~they tend to be oriented along the crystallographic lattice of graphene.\\

Finally, we wish to extend the discussion beyond SiO$_2$ as a substrate because folding by SHI can be achieved in graphene supported on virtually any substrate. From the huge difference in size and length for supported and suspended graphene (see Fig.~\ref{1}) one can already conclude that the substrate does play an important role for the folding. Furthermore, whenever graphene is exfoliated or transferred onto its supporting substrate material, an interfacial layer will be present after preparation. This layer will consist mostly of water and will significantly influence the physical properties of graphene. For example, it can lead to blocking of charge transfer from the substrate to graphene \cite{Ochedowski.2011,Durso.2012,Shim.2012,Ochedowskiii.2014} or reduce the adhesion forces between graphene and the substrate \cite{Gao.2011}. Upon irradiation the SHI will interact with the substrate and the interfacial layer along its trajectory and the intensive and abrupt heating of the substrate material as well as the trapped water layer can lead to a significant ejection of particles which will certainly affect the folding process. 

\begin{figure}[ht]
\centering
\includegraphics[width=15cm,keepaspectratio]{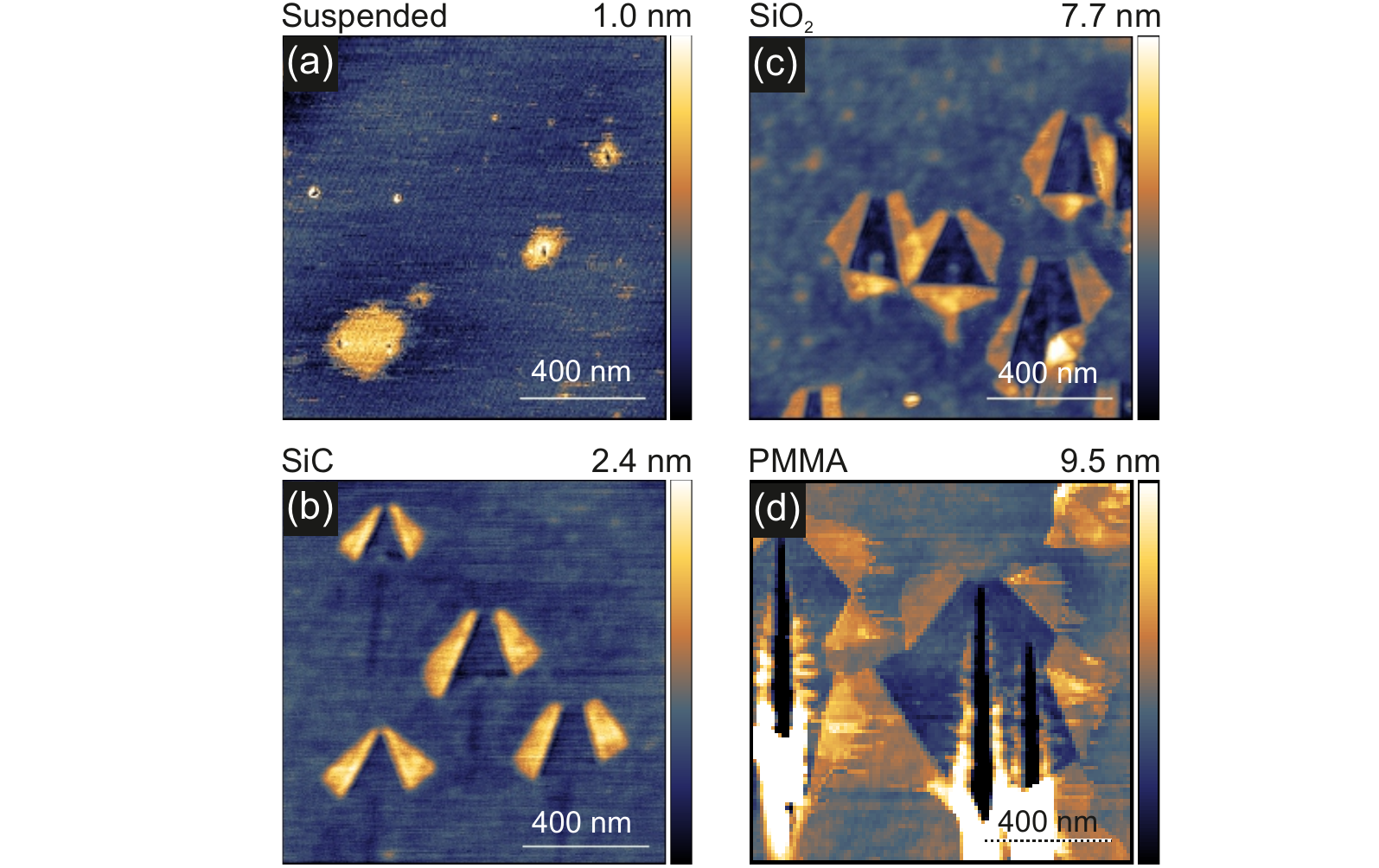}
\caption{AFM images of graphene foldings fabricated by SHI irradiation ($dE/dx=15$~keV/nm, $\Theta$=1.5$\pm0.3^{\circ}$) on different substrates. For better comparison all images are shown in the same size but with different height scales shown to the left of each image. (a) Suspended graphene, (b) graphene on SiO$_2$, (c) graphene on SiC, and (d) graphene on PMMA.}
\label{5}
\end{figure}

To demonstrate the role of the substrate more clearly, we have compiled data obtained from various substrate supported graphene samples as exemplary shown in Fig.~\ref{5}. Here, the incidence angle has been kept constant at $\Theta=1.5\pm0.3^{\circ}$. In suspended graphene (Fig.~\ref{5}(a)), the irradiation does create defects which appear as holes rather than foldings in these images. However, as we have just shown, they are indeed foldings and for the sake of clarity we will stick to the term. With a typical length of 20 -- 30~nm (averaging over 10 foldings) the foldings in suspended graphene are much shorter than on any of the substrates. Graphene on SiO$_{2}$ (Fig.~\ref{5}(b)) shows foldings with a length of 150~nm on average which is about five times the value of suspended graphene. Note, that stopping power and angle of incidence were kept constant here, but that the different substrates react quite differently to the irradiation. The SiO$_{2}$ substrate shows faint surface tracks after SHI irradiation, while SiC just shows very shallow depressions because only the topmost Si atoms are sublimating upon irradiation~\cite{Ochedowski.2014}. Nevertheless, the foldings in graphene on SiC (Fig.~\ref{5}(c)) are almost of the same length as in graphene on SiO$_{2}$. With the radiation sensitive polymer PMMA as substrate (Fig.~\ref{5}(d)), the foldings are more than twice as long (around 370~nm) as the foldings on SiO$_{2}$ and their length exceeds the length of the foldings in freestanding graphene by as much as a factor of 15. In addition, the foldings are also broader. This nicely demonstrates that apart from the interfacial water layer, the amount of material ejected from the substrate due to the interaction with the ion influences the size of the folding to a large extent. For the most common conditions used in current basic research, i.e.~a SiO$_2$ substrate with a water layer present, very good results with respect to folding size and efficiency are easily obtained. In freestanding graphene, where both substrate and interfacial layer are absent, nanopores with a typical width of 10 -- 20~nm can be fabricated. 

This, together with the fact that the shape of the nanopore can be tailored by various parameters proves that SHI are indeed a powerful tool to perforate supported and free-standing graphene. Nanoporous graphene has been suggested as a key factor in many applications. For example, graphene nanomeshes supported on SiO$_2$ have recently been shown to be semiconducting thus offering new possibilities for bio-sensors and spintronic devices \cite{Bai.2010}. Graphene perforated by SHI may not yield the inherent periodicity of nanomeshes but as an advantage the pore density is much easier to control. Swift heavy ions represent the standard technique to fabricate track etched membranes \cite{Trautmann.2009} and based on our results we propose that the concept can be extended to graphene-based membranes which promise an unprecedented level of transport rates (permeability) in filtering applications as the quasi two-dimensional selective membrane would exhibit negligible wall interactions \cite{Liu.2014}. 
\\
\\
\noindent
{\bf Summary}\\
In summary, we could show that graphene is very sensitive to ionizing particle radiation. As a consequence, swift heavy ion beams can be used as a versatile tool to intentionally fold graphene, yielding nanopores and graphene nanostructures with specific characteristics such as size, orientation or shape. In general, the technique offers a high level of control as ion beam parameters are easily changed, controlled and maintained. To achieve perforation and folding of graphene, there are two conditions to be met. First, the ion has to deposit sufficient energy per track length to induce vacancy like defects in the graphene sheet ($S_{e}\geq$ 5~keV/nm). This can be readily achieved with small scale accelerators as they are commonly used in standard ion beam facilities. Second, graphene has to be irradiated under non-perpendicular angles of incidence. This poses no severe limitation as in supported graphene on SiO$_2$ e.g., foldings occur already at angles as large as $\Theta\simeq$~45$^{\circ}$. However, for tailoring specific nanostructures, additional criteria have to be fulfilled. While the length of the folding are mainly determined by the incidence angle $\Theta$ and the energy loss S$_{e}$, our comprehensive experimental studies show that width and shape of the folding can be tuned in a wide regime by the appropriate choice of the substrate material. 
Of all two-dimensional materials graphene will probably be the one which is the least susceptible to effects of ionizing irradiation. Therefore, we are confident that swift heavy ions are a useful tool to fabricate similar nanostructures in other 2D materials such as MoS$_2$ and hexagonal BN, or even heterostructures thereof.

\section{Methods}
\label{Exp}
\noindent
Supported graphene samples have been prepared by mechanical exfoliation of a HOPG crystal (Momentive Performance) onto a standard 90~nm SiO$_2/$Si (Graphene Supermarket) wafer under ambient conditions. 
Graphene was grown on 25 $\mu$m thin copper foils (99,8\%, Alfa Aesar GmbH \& Co KG, Karlsruhe) using low-pressure chemical vapour deposition \cite{Cai.2009,Woszczyna.2014}. Before growth the copper foils were rinsed with acetone then placed in acetic acid (p.a.) for 10~min and finally cleaned with water, acetone and isopropanol. Then the foils were dried under a nitrogen stream, put into a quartz tube and introduced into a furnace (Gero F40-200, Neuhausen). Pumping the tube to $1\times 10^3$~mbar was followed by setting a H$_2$-flow of 50~sccm and subsequent heating to 1288~K in two steps (500K/h up to 1223~K, 100K/h up to 1288~K). After 3~h the H$_2$-flow was reduced to 10~sccm and a methane flow of 70~sccm was introduced for 15~min. Then the samples were cooled down to room temperature under the Ar/H$_2$ atmosphere. The grown graphene sheets were transferred onto TEM grids. To this end, poly(methyl methacrylate) was spun in two steps onto the graphene to stabilize it during the transfer process. First, a layer of low molecular weight PMMA (50k) and then a second layer of high molecular weight PMMA (950k) was spin-cast for 30~s at 4000~rpm onto the graphene. Then the back site of the copper foil was cleaned using O$_2$ plasma etching for 60 s, followed by etching away the copper using an ammonium persulfate solution (6.5~g ammonium persulfate/100~ml water) over night. After placing the graphene onto the Quantifoil TEM grid, the PMMA was dissolved in a critical point dryer. To additionally clean graphene the samples were introduced into ultra-high vacuum ($4 \times 10^{-9}$~mbar) and annealed for 2~h at 350~\DegreeC.

Samples haven been transferred to the different accelerators for irradiation. In the high energy regime ($\geq 0.5$~MeV/u), ion irradiation leads to ionization and excitation of target atoms. The most important parameter to achieve a given modification is thus the electronic stopping power $dE/dx$. This deposited energy per track length depends on the projectiles mass, its kinetic energy, and the target properties (mass, stoichiometry, density). Thus, with a given sample material, the desired stopping power can be achieved by simply tuning the kinetic energy of the projectile. This again requires accelerators which operate with many different ions at the various beam times. In the case of graphene, a given modification will be achieved as long as a certain threshold is overcome (see main text) independent on the type of ion that was used for the respective irradiation. For simplicity and easier comparison we have therefore calculated  the stopping power in graphite with the standard software code SRIM~\cite{Ziegler.2010} and used these numbers throughout the manuscript instead of naming the specific ion type and its energy. Samples have been irradiated with various ions (see table below) at the IRRSUD beamline of the GANIL (Caen, France) and at the Tandem Van de Graaff accelerator at the RBI (Zagreb, Croatia) which are both equipped with a positioning system for glancing incidence irradiation. Irradiation took place under vacuum conditions with $p_{base}\leq10^{-6}$~mbar. 
\\\\
\noindent
\begin{tabular}[htb]{|c|c|c|c|}
\hline
Swift heavy ion  & $S_e$ keV/nm & $S_n$ keV/nm & Range $\mu$m\\
\hline
\hline
3 MeV O   & 1.9  & 0.01 & 2.5\\
6 MeV Si  & 3.3  & 0.02 & 3.0 \\
15 MeV Si & 4.3 & 0.01 & 5.3 \\
23 MeV I  & 6.0  & 0.27 & 6.2\\
84 MeV Ta & 14.0   & 0.23 & 11.1\\
91 MeV Xe & 14.8 & 0.10 & 11.4 \\
106 MeV U & 16.1 & 0.36 & 11.8 \\
\hline
\end{tabular}
\\
\\
\noindent After irradiation samples have been analyzed by AFM, Raman and AC-HRTEM. The AFM measurements were performed using a Veeco Dimension 3100 AFM in tapping mode. The cantilevers used for these measurements are Nanosensors NCHR with a typical resonant frequency of 300~kHz. As absolute height measurements of 2D materials in tapping mode are difficult to obtain, a typical height scale for our samples is shown only in Fig.~\ref{1} and in Fig.~\ref{5}, where large variations due to the substrate modifications occur. For the rest of the AFM images the height scale is given in the figure captions instead. Raman spectra were taken with a Renishaw InVia Raman microscope. The laser excitation wave length was 532~nm and the laser intensity was always kept below 0.4~mW with a spot size below 1~$\mu$m to avoid heating effects. \\
\\
The high-resolution transmission electron microscopy was conducted using an FEI Titan 80-300 microscope equipped with an image side hexapole spherical aberration corrector and a field emission gun. The extraction voltage of the gun was lowered to 2~kV to reduce the energy spread of the beam. The
spherical aberration was tuned to $\approx 20~\mu$m and the imaging was conducted at Scherzer focus, resulting in dark atom contrast.

\section{Acknowledgement}
Financial support from the DFG within the SPP 1459 "Graphene" is gratefully acknowledged. MK and MJ acknowledge the Croatian Science Foundation (pr.no. 8127) which supported this investigation. Support by the Croatian Centre of Excellence for Advanced Materials and Sensing Devices is also acknowledged. We thank V.~Freie Soler for support with sample preparation.

\section{References}

\end{document}